\documentclass[lettersize,journal]{IEEEtran}
\usepackage{amsmath,amsfonts}
\usepackage{algorithmic}
\usepackage{array}
\usepackage[caption=false,font=normalsize,labelfont=sf,textfont=sf]{subfig}
\usepackage{colortbl}

\usepackage{textcomp}
\usepackage{stfloats}
\usepackage{url}
\usepackage{verbatim}
\usepackage{graphicx}
\def\BibTeX{{\rm B\kern-.05em{\sc i\kern-.025em b}\kern-.08em
    T\kern-.1667em\lower.7ex\hbox{E}\kern-.125emX}}
\usepackage{balance}
\begin{document}
\title{AdaptNet: Rethinking Sensing and Communication for a Seamless Internet of Drones Experience
}
\author{Ananya~Hazarika,~\IEEEmembership{Student Member,~IEEE} and
        Mehdi~Rahmati,~\IEEEmembership{Senior Member,~IEEE,}
        
\thanks{\IEEEcompsocthanksitem A.~Hazarika and M.~Rahmati are with the Department
of Electrical and Computer Engineering, Cleveland State University, OH, 44115 USA \protect 
E-mails: 
a.hazarika@vikes.csuohio.edu and m.rahmati@csuohio.edu. \protect
}
}

%tempfor arxiv: 
\markboth{\textbf{This is the author's version of the paper, accepted for publication in the IEEE Internet of Things Magazine, July 2024. }}{}
%\markboth{IEEE Internet of Things Magazine,~Vol.~xx, No.~x, August~2024}%
%{Dynamicity-Aware Joint Sensing and Communications in the Internet of Drones}

\maketitle

\begin{abstract}
In the evolving era of Unmanned Aerial Vehicles~(UAVs), the emphasis has moved from mere data collection to strategically obtaining timely and relevant data within the Internet of Drones~(IoDs) ecosystem. However, the unpredictable conditions in dynamic IoDs pose safety challenges for drones. Addressing this, our approach introduces a multi-UAV framework using spatial-temporal clustering and the Fréchet distance for enhancing reliability. Seamlessly coupled with Integrated Sensing and Communication~(ISAC), it enhances the precision and agility of UAV networks. Our Multi-Agent Reinforcement Learning~(MARL) mechanism ensures UAVs adapt strategies through ongoing environmental interactions and enhancing intelligent sensing. This focus ensures operational safety and efficiency, considering data capture and transmission viability. By evaluating the relevance of the sensed information, we can communicate only the most crucial data variations beyond a set threshold and optimize bandwidth usage. Our methodology transforms the UAV domain, transitioning drones from data gatherers to adept information orchestrators, establishing a benchmark for efficiency and adaptability in modern aerial systems.
\end{abstract}


\section{Introduction}
Beyond-5G~(B5G)/6G networks are poised to redefine the digital communications landscape, offering unparalleled speed, reliability, and ultra-low latency, surpassing the achievements of the 5G New Radio~(NR). Considering the criticality of modern digital landscape, heterogeneous networks are modeled to efficiently integrate the prevailing heterogeneities such as communication modalities, channel types, and Quality of Service~(QoS) requirements~\cite{singhal2017resource}. With recent technological advances, the concept of the Internet of Things~(IoT) is maturing into an omnipresent digital mesh, influencing nearly every aspect of human life and industrial operations. The emerging Internet of Drones~(IoD) extends the capabilities of traditional IoTs into the aerial domain, with next-generation B5G/6G networks promising adaptive, interconnected frameworks for ubiquitous coverage, enhanced intelligence, and superior safety~\cite{mishra2021drone}. This convergence represents a significant shift towards integrating communication and sensing, transforming networks into \textit{digital sixth sense} that enhances human cognition. This evolution enables drones to utilize real-time sensory data for navigation and intelligent decision-making, marking a transition from traditional  roles to dynamic entities that provide critical insights into the physical environment.
However, as these advanced networks prepare to support the escalating demands of IoDs, they encounter significant challenges. These include managing increased network complexity, ensuring seamless connectivity in high-mobility scenarios, and addressing the intricate balance between rapid data transmission and energy-efficient operations in UAV networks. One of the foremost challenges in contemporary UAV networks is the limited bandwidth availability, which restricts data transmission rates and impacts the efficiency of UAV operations. Additionally, UAVs are often limited by battery life, posing a significant challenge for long-duration missions and continuous data transmission. Real-time data processing and decision-making are critical in UAV networks, yet high latency and limited onboard processing power often hinder timely responses to dynamic environmental changes~\cite{li2018uav}. Additionally, ensuring precise navigation and target detection in cluttered or dynamically changing environments remains a significant challenge for UAVs. 
\begin{figure}[!t]
\centering
\includegraphics[width=0.95\columnwidth]{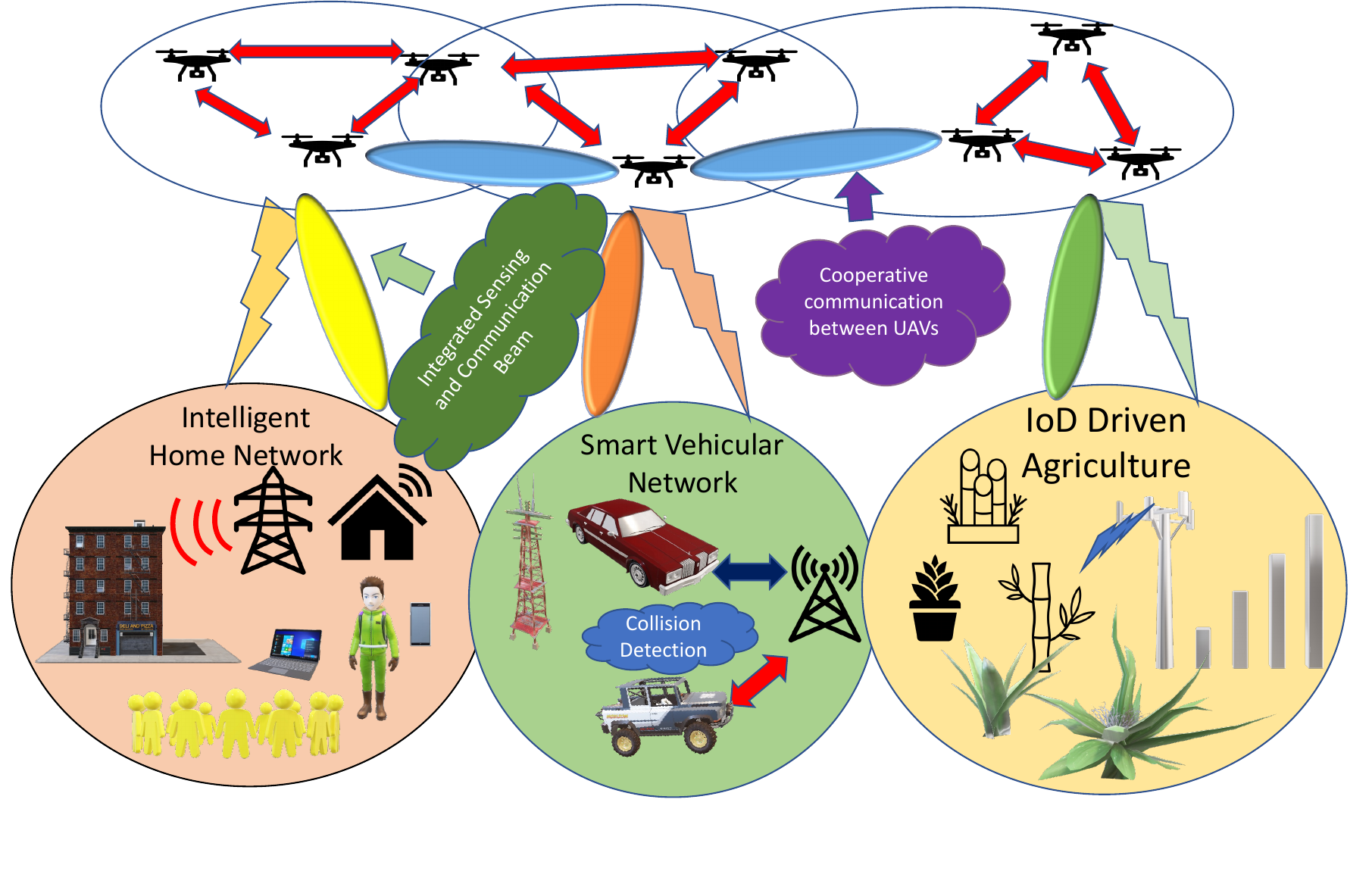}
\caption{The broad spectrum of use cases for UAVs employing ISAC.}\label{usecase1}
\vspace{-2mm}
\end{figure}

Considering the above-mentioned challenges, the merging of IoDs with dynamic multi-class targets requires adaptive communication protocols due to the unpredictable behaviors of these targets. These protocols ensure that the targets receive the information they need, thus preventing redundancy and conflicts in high-stakes scenarios.
As 6G facilitates extensive data exchange and AI applications, ensuring the safe and efficient operation of UAVs in this environment becomes crucial.

In response to these intricate challenges, we present AdaptNet, a transformative architecture aiming to redefine the dynamics of data acquisition and transmission in an IoD network. This architecture, inspired by the Integrated Sensing and Communication~(ISAC) model~\cite{ahmadipour2022information}, is purpose-built to cater to multi-class targets by seamlessly integrating adaptive MIMO radars and the precision of the Fréchet distance for optimal target detection. We utilize ISAC, an emerging B5G/6G technology capable of integrating sensing and communication functions into a unified framework, mainly used in the context of UAVs. Figure~\ref{usecase1} demonstrates the diverse applications of UAVs equipped with ISAC technology, showcasing its utility in smart home networks for enhancing household automation and connectivity, in vehicular networks for real-time traffic management and vehicle-to-vehicle communications, and in smart agriculture for efficient crop health monitoring and precision farming. The Fréchet distance~\cite{har2014frechet} is an advanced metric for assessing similarity between curves, considering both the spatial positions and the sequential order of points. It is particularly beneficial for analyzing complex data sets such as time-series, trajectories, and multidimensional sequences, where the arrangement and ordering of data points are key factors. By embedding the Fréchet distance into our system, we gain a powerful tool for identifying subtle yet significant shifts in environmental data patterns critical for the real-time decision-making processes in ISAC-enabled UAV systems. This unique blend ensures UAVs navigate precisely, capturing relevant data and adapting to real-time environmental challenges. By leveraging radar's detailed scanning capabilities combined with the evaluative power of the Fréchet distance, AdaptNet guarantees efficient resource allocation and Ultra-Reliable, Low-Latency Communications~(URLLC). Through this approach, drones transition from essential data collectors to proactive managers of information, marking a significant advancement in UAV network efficiency and adaptability.

In our exploration to optimize UAV network operations, we have seamlessly integrated Multi-Agent Reinforcement Learning (MARL)~\cite{wen2022multi} with the ISAC paradigm. This integration allows multiple UAVs, each acting as an independent agent within the network, to learn and adapt their sensing and communication strategies in real-time. The agents interact with and learn from the environment and the actions and experiences of other agents in the network. This collaborative learning process leads to more efficient and effective overall system performance as UAVs continuously refine their strategies to meet the dynamic demands of their operational tasks. 
However, in scenarios demanding robust data exchange among UAVs, our focus pivots to communications, where UAVs, acting as individual agents, apply the Multi-Agent Deep Deterministic Policy Gradient~(MADDPG) algorithm. MADDPG, known for its advanced learning mechanism~\cite{fan2021multi} that combines deep learning with deterministic policy gradients, facilitates a centralized learning and decentralized execution approach. This optimizes the timeliness and relevance of data transmission, with UAVs making instant optimal decisions by continually learning and adapting from their environment and the strategies of other agents. This robust approach enables UAVs to navigate the complex dynamics of multi-agent interactions, ensuring efficient and coordinated communication within the network.
{AdaptNet's practical applications extend through various sectors, significantly improving UAV functionalities within smart home environments, vehicular networks, and agricultural operations. It enhances household automation and connectivity, integrating seamlessly with IoT devices for dynamic home management. It can also facilitate real-time traffic management and vehicle-to-vehicle communication in vehicular networks, improving road safety and efficiency. Through this novel architecture, drones optimize crop health monitoring and management, enabling precise agriculture with data-driven insights. This integration enables UAVs to adapt quickly, ensuring efficient data communication and advanced sensing capabilities. By transforming theoretical concepts into real-world applications, AdaptNet sets a new standard for adaptability, efficiency, and precision in UAV network management across diverse environments.}

Based on the mentioned requirements, our work presents these key advancements:
{\begin{itemize}
\item AdaptNet integrates the sensing with Fréchet distance precision for elevating UAV target detection and navigation, transforming drones from basic data collectors into strategic information managers.
\item Through utilizing MARL with ISAC, AdaptNet empowers UAVs to adapt strategies dynamically. This approach ensures agile and secure operations in dynamic environments, enhancing real-time decision-making and responsiveness.
\item AdaptNet employs the MADDPG algorithm, refining UAV communication policies for enabling efficient, timely, and relevant data transmission and setting new performance standards for UAV networks while addressing challenges such as bandwidth limitations, energy management, and latency reduction.
\end{itemize}}
\section{Our Approach}
\begin{figure*}[!t]
\centering
\includegraphics[width=5.2in]{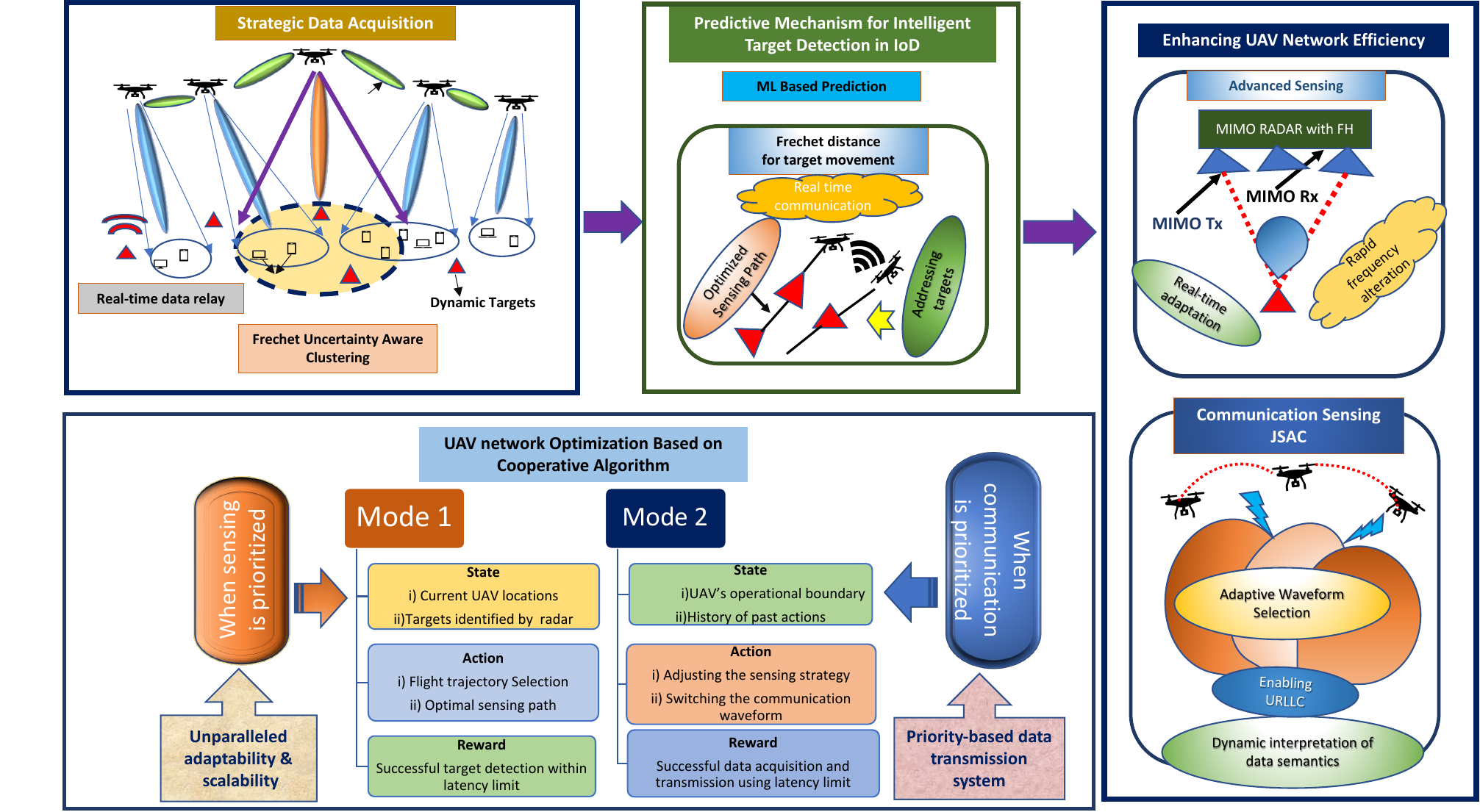}
%\vspace{-2mm}
\caption{Visual Overview of AdaptNet Architecture and its workflow from intelligent data acquisition in the IoD to dynamic cluster prediction utilizing the Fréchet distance, and to cooperative algorithm enhancements for improved UAV network efficiency and adaptability. It includes detailed segments depicting the coordination of state, action, and feedback mechanisms within MARL and MADDPG frameworks, aimed at prioritizing sensing and communication. Furthermore, it highlights AdaptNet's strength in prioritizing data relevance through adaptive waveform selection, showcasing its effectiveness across various network operations}\label{usecase_fig}
\end{figure*}

As depicted in Fig.~\ref{usecase_fig}, the key to AdaptNet's architecture is set to enhance UAV sensing and communications through aggregated higher data rates and reduced latency. This integration will be pivotal in real-time, critical scenarios such as disaster response and environmental monitoring, leveraging AdaptNet's advanced adaptive capabilities.
By enabling drones to swiftly adjust their strategies in response to changing target movements and environmental conditions, AdaptNet minimizes delays and maintains consistent communications. This approach boosts IoD's operational safety and ensures efficient and risk-minimized data transfer in dynamic settings.

\subsection{ISAC for Fast-moving Targets} 
Given the ever-changing trajectories of targets and the inherent high mobility of drones, they can be error-prone to inaccuracies. ISAC framework can address these challenges by ensuring seamless synchronization between sensing and data transmission. Our approach integrates the strengths of sensing and communication under one umbrella, prioritizing the real-time communication of sensed data, thus enhancing the adaptability and responsiveness of drones to dynamic environments. Such a focus on communication ensures that the drones not only gather accurate data but are also swift in disseminating this information, reducing latency and making real-time adaptations possible. However, environmental factors can significantly disrupt drone operations, leading to deviations that, when coupled with uncertainties, might result in missed targets or, in the most severe cases, accidents~\cite{pitre2012uav}. Our strategy utilizes the power of Machine Learning~(ML) to bolster the ISAC framework, enhancing the accuracy and utility of target information collected by IoDs. 

\subsubsection{Enhancing IoD Efficiency through Fréchet Distance-Based Sensing}
We propose using adaptive radars to ensure consistent, real-time data in dynamic environments and strengthen the foundation of URLLC. Addressing the challenges of multi-target sensing in UAV networks~\cite{liu2023robust}, our solution integrates the precise spatial-temporal clustering of the Fréchet distance~\cite{har2014frechet} with the capabilities of radar-centric ISAC.
 The Fréchet distance provides a robust metric for assessing the similarity between various data trajectories, factoring in their spatial positions and temporal sequences. This analysis allows UAVs to identify patterns and changes over time with greater precision, facilitating the efficient clustering of targets based on their movement and behavior patterns.
 Using this combination, UAVs can identify the best routes for their missions, pinpoint target positions, and schedule sensing accordingly.
A smaller Fréchet distance indicates paths are more alike, directing smart UAV resource use and ensuring URLLC. 
By incorporating a radar-centric ISAC framework into our architecture, we capitalize on radars with real-time, extensive range capabilities, finely tuned for capturing the dynamic nature of UAV networks.  
By leveraging radar's innate capabilities, we empower our system to identify the dynamic characteristics of targets within IoD networks. The radar units, working with the ISAC approach, constantly scan the environment, capturing intricate details of targets' movement and interactions.
Our methodology extends beyond the fundamental use of radar for sensing; it leverages its potential to unravel the complex dynamics of movements within UAV networks. Hence, this integration, coupled with the Fréchet distance as a guiding metric, provides a holistic view of the environment's evolving dynamics. 

MIMO radars, enhanced with Frequency Hopping (FH)~\cite{9656537}, offer a unique adaptability critical in the dynamic world of IoDs. We specifically consider the deployment of pulsed fast FH, where signals are continuously emitted with rapidly alternating frequencies multiple times within a Pulse Repetition Interval (PRI), followed by a silent period~\cite{paisana2017signal}. The specific deployment of FH radar within this paradigm improves the radar-centric ISAC by diversifying and rapidly alternating the transmission frequencies. 
As drones interact in increasingly crowded skies, the IoD necessitates quick, reliable, and secure communication. The adoption of FH radar addresses this challenge by mitigating the need for expensive instantaneous wideband components while maintaining superior sensing performance. This inherent ability of FH radar to quickly hop between frequencies can be strategically deployed to reduce the chances of communication collision within the IoD, making real-time communication more robust and efficient.
 Hence, this fusion of MIMO and FH radar along with the application of the Fréchet distance ensures that our drones in the IoD ecosystem navigate with unmatched precision, perfectly synchronized to the demanding rhythms of today's skies.
\subsubsection{Prioritizing Data Relevance for Optimized UAV Communications}
In our research forefront, the emphasis begins with the importance of advanced radar systems in dynamically acquiring sensing data from uncertainty-aware clustered regions. In the dynamic IoD network, the initial challenge lies in gathering vast amounts of data by intelligently identifying and capturing the most pertinent data in real-time. However, traditional radar techniques often stumble when faced with unpredictable environments or erratically moving targets. To counter these challenges, we employ the proposed radar system adept at acquiring and analyzing data's relevance, with the Fréchet distance serving as a vital tool in this assessment. Our novel process forms a crucial part of our framework for real-time data relevance assessment, which is instrumental in determining the novelty and relevance of the new information. By adopting the methodology outlined below, we ensure the UAV network is susceptible to the importance of the data, enabling effective and agile communications to adapt to changing environmental conditions.

{\textbf{Fréchet Distance Analysis:} When the UAV's radar system captures new data, it computes the Fréchet distance between this data and existing data patterns. This calculation compares raw data points and a comprehensive analysis considering the data's geometric patterns and inherent trends. The Fréchet distance effectively encapsulates the overall shape and trajectory of the data sets, offering a robust similarity measure. A low Fréchet distance implies a high degree of similarity, indicating potential redundancy, while a high distance suggests significant changes or novel information in the sensed environment.}

{\textbf{Threshold-Based Decision Mechanism:} Our system employs a predefined threshold for the Fréchet distance. If the calculated distance falls below this threshold, it suggests that the newly sensed data is closely aligned with existing information, deeming it less critical for immediate transmission. This approach aids in bandwidth and resource conservation by avoiding the transmission of redundant information. Conversely, a Fréchet distance exceeding the threshold highlights the emergence of new, potentially critical environmental changes. In such instances, the system prioritizes the communication of this data, recognizing its importance for real-time decision-making and actions.}

{\textbf{Integration with MU-MIMO Communications: }}  We propose the deployment of the Multiuser MIMO~(MU-MIMO) communications waveform as its uniqueness lies in its adaptability, steered by insights derived from the radar signals via the Fréchet distance metric. Essentially, the Fréchet distance becomes our barometer, determining the significance and urgency of the data sensed by the UAVs. When this distance surges, highlighting a growing disparity or importance in the sensed data, our system nudges the UAV to activate a high-throughput waveform. Even though this transmission mode demands more power, it's a trade-off for ensuring that vital data is relayed immediately. If the data from the radar is similar to past information shown by a lower Fréchet distance, then the UAV can choose a more energy-saving waveform while fetching relevant data.
This intelligent oscillation between waveforms, steered by the radar's insights and the Fréchet metric, equips our UAV network to maintain an optimal balance. Such real-time adaptability in communications ensures that UAVs are both energy-efficient and agile in their response to evolving scenarios. 
\subsection{Communications Policy for Uncertainty Mitigation}
With increased data volume gathered in an ISAC-based IoD system, traditional communications protocols can struggle to maintain efficiency. Hence, the system must prioritize decisions based on which data needs immediate transmission and which can be delayed, compressed, or ignored. We consider Age of Information (AoI)~\cite{kosta2017age} as a quantitative benchmark for gauging the timeliness of packet reception, thus defining the freshness of information. 
With AoI as a metric, we can evaluate the promptness of data shared among the UAVs. Ensuring real-time or near-real-time updates is vital for both safety and operational purposes. To achieve this, we explore three queuing models: First-Come-First-Served (FCFS), Last-Come-First-Served with preemption in service (LCFS-S), and Last-Come-First-Served with preemption allowed only while waiting (LCFS-W)~\cite{hazarika2022framework}. These models give insights into how data packets, including sensing and communications information, are processed and relayed, ensuring that the most recent data is always prioritized.

After acquiring data through the Fréchet distance metric, we seamlessly transmit this information into the proposed communication infrastructure. This synergistic integration goes beyond merely enhancing the UAVs' sensing trajectories; it ensures that the communication framework remains responsive, adaptive, and primed to manage fluctuating data magnitudes. {Significantly, this integration establishes the foundation for our innovative prioritization strategy, i.e., data deemed highly relevant, identified through deviations measured by the Frechet distance, receives priority for immediate transmission. Meanwhile, less urgent data is methodically organized into a queue for subsequent transmission.} Such a strategy ensures that UAVs within the network consistently receive timely and pertinent information.
\section{The Proposed Cooperative Algorithm}
In UAV environments, the need to adeptly manage both sensing and communication tasks under constantly changing conditions, like moving targets and fluctuating communication channels, necessitates distinct operational strategies. Our approach introduces two modes, with Mode 1 prioritizing advanced radar-based sensing and Mode 2 focusing on optimizing communication. These modes, driven by a feedback loop, significantly enhance the system's adaptability. By continuously monitoring feedback on the relevance and quality of received data, UAVs dynamically adjust their sensing and communication tactics, whether to prioritize immediate environmental awareness (Mode~1) or efficient data transmission (Mode~2). If patterns suggest frequent redundant or less significant data, the UAVs recalibrate to ensure efficient resource usage, always prioritizing the most vital data. This continuous feedback and adjustment process is reminiscent of reinforcement learning, where systems learn and adapt based on feedback from their actions. With each UAV acting as an independent agent, its actions can affect the learning landscape for others, making it a complex environment to navigate. 

\subsection{Mode~1. When Sensing Is Prioritized}

In Mode~1, our primary focus is efficiently managing radar-based sensing in dynamically changing scenarios. The key challenge here is acquiring data and ensuring that the sensing is adaptive and responsive to the changing environment. Here, the IoD network employs MARL to enhance radar sensing capabilities. Each UAV acts as an agent, collaboratively learning and adjusting its sensing strategies based on collective intelligence. This mode is particularly crucial when the primary objective is to maintain up-to-date environmental awareness through advanced sensing techniques.

The importance of Mode~1 lies in its ability to navigate the complexities of radar sensing in an ever-changing operational landscape. Considering the complexities of balancing radar-sensing across numerous UAVs, our approach utilizes MARL to transition from individual optimization to a sophisticated orchestration of UAV interactions. 
It is not just about a single UAV adjusting to its radar-collected data but an ensemble of UAVs collaboratively responding to collective intelligence and adapting in rhythm.

Our proposed state space in the MARL framework amalgamates various elements. {As depicted in Fig.~\ref{usecase_fig}, the state space is shown which includes the UAV's current positions and trajectories and the targets identified by radar and their evolving dynamics. The action domain for UAVs is shown to be multidimensional as it deals with choices about flight trajectories, determining the radar sensing path that ensures the quickest data acquisition. The reward mechanism is both intelligent and comprehensive, as successful target detection within set latency benchmarks earns positive reinforcements.} Using the cooperative approach of MARL, our UAVs work together in their training, with a unified focus on common objectives, such as reducing latency or enhancing the precision of target detection. Their shared experiences and collective rewards provide valuable insights, guiding UAVs to select optimal routes, improve target detection, and align sensing tasks seamlessly. {The outcome of this mode is a UAV network characterized by unmatched adaptability, scalability, and operational efficiency. This integration facilitates URLLC and refines UAV navigational paths and sensing methods to respond to dynamic environments adeptly.}
\subsection{Mode~2. When Communication Is Prioriized} {In the complex landscape of IoD networks utilizing ISAC, there are scenarios where communication takes precedence, particularly when the need for rapid data transmission arises. Considering this requirement, each UAV in Mode~2 is a primary aerial vehicle, evolving into an autonomous agent with operational parameters defined within a specific state space. This mode activates when the efficient transmission of collected data becomes more critical than gathering new sensory inputs. In this context, the UAVs leverage the Multi-Agent Deep Deterministic Policy Gradient (MADDPG) algorithm~\cite{li2019robust} to optimize their communication strategies intelligently.}

 As shown in Fig.~\ref{usecase_fig}, the state space for each UAV agent includes its operational boundary information and historical actions, offering insights into the impact of past decisions. Agents within this framework make strategic choices from their action space that involve adjusting sensing strategies, modulating communication waveforms, and prioritizing data transmission. The reward function, pivotal to the learning process, allocates positive reinforcement for successful, timely data transmission and penalizes actions leading to redundancy, excessive energy use, or communication breakdowns. Through the MADDPG algorithm, UAVs in this mode collaboratively learn to adapt their actions based on individual observations and other agents' collective experiences. This cooperative learning mechanism minimizes latency, maximizes the relevance of transmitted data, and ensures judicious use of resources. As UAVs encounter diverse scenarios, they continuously refine their policies to enhance performance, balancing the need for immediate data dissemination, energy conservation, and bandwidth conservation.

MADDPG uses deep neural networks to approximate the Q-functions and policies for each agent. It employs an off-policy approach where agents learn a deterministic policy that maps their state to an action. The algorithm trains by optimizing the Q-function~\cite{lowe2017multi}, minimizing the difference between the estimated and actual rewards. The unique aspect of MADDPG is its centralized learning but decentralized execution approach. During training, agents have access to the observations and actions of all other agents, allowing them to learn the best policy considering the actions of others. However, during execution, each agent uses its policy to select actions based solely on observation. The agents are trained collaboratively, sharing experiences and global reward signals to optimize their sensing and communication strategies. {This innovative approach elegantly balances resource efficiency with the imperative of delivering crucial data, setting new benchmarks in UAV-driven sensor network capabilities.}

\section{Results and Discussion}
\begin{figure*}[!t]
\centering
\includegraphics[width=6.1in]{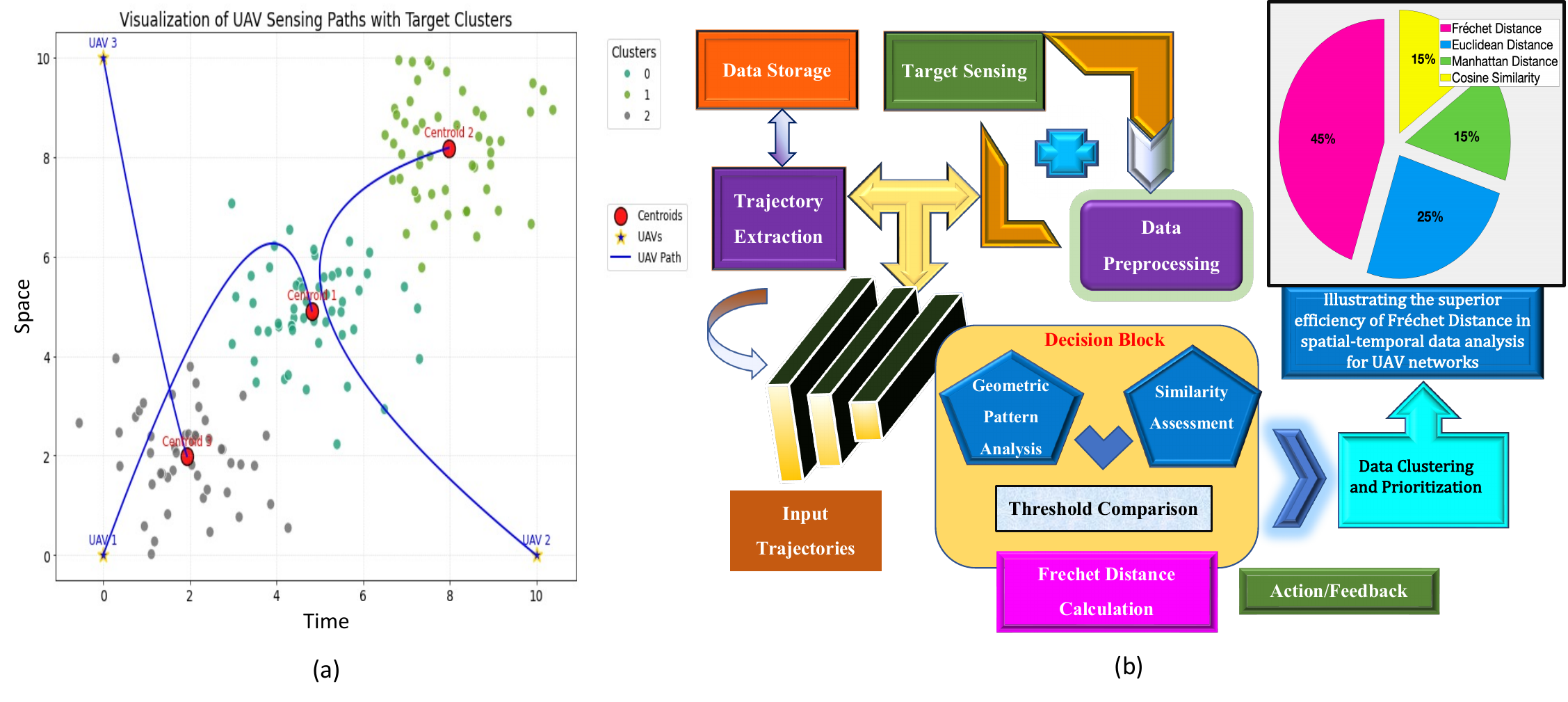} 
\vspace{-0.2cm}
\caption{(a) Optimal sensing path of the UAVs along with the clustered fast-moving targets using Frechet distance; {(b) Operational flow of Frechet distance analysis in the IoD framework. This block diagram shows the sequential process from initial target sensing to the final computation of Frechet distance. It starts with target detection and advances through preprocessing for trajectory input formulation. Subsequently, it undertakes geometric pattern analysis of these trajectories, engages in threshold comparisons for pattern similarity, and performs the precise calculation of the Frechet distance for effective target clustering and prioritization.} }
\label{fig:frecblock}
\end{figure*}

\begin{figure}[!t]
    \centering
   \includegraphics[width=0.90\columnwidth]{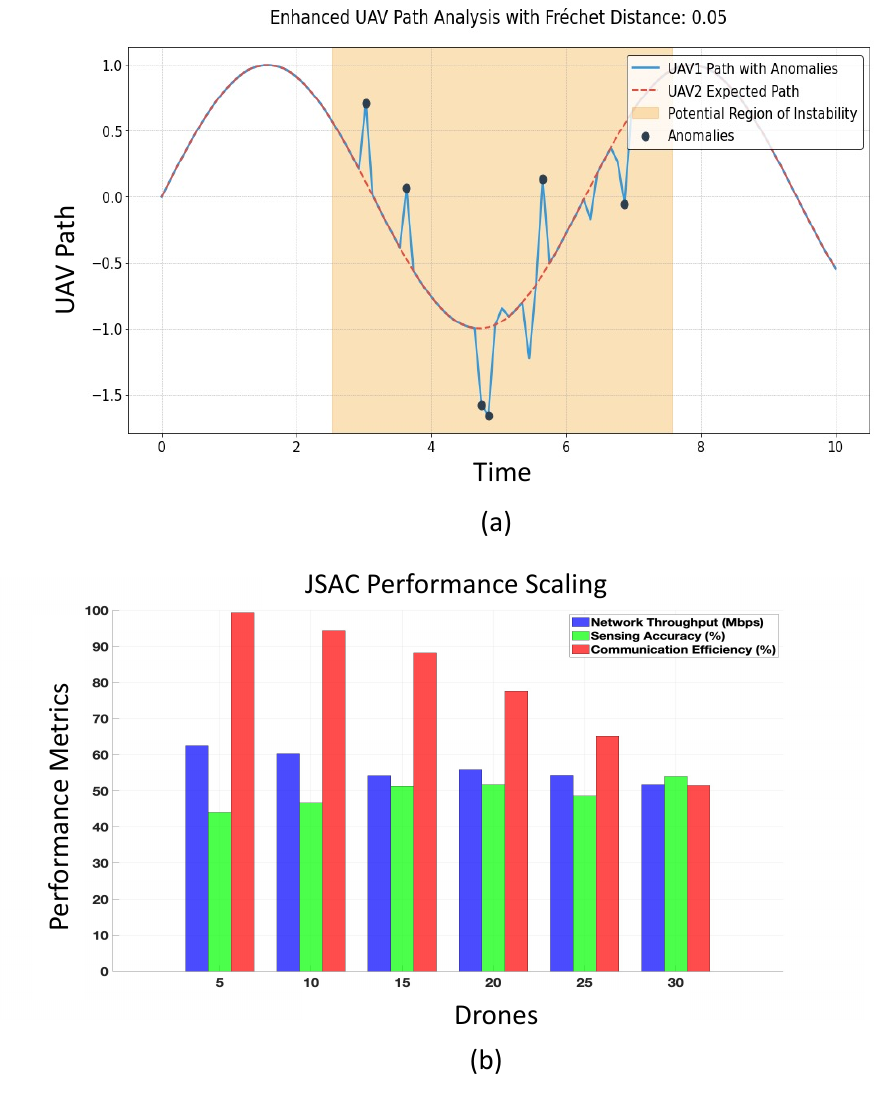}
    \caption{(a) A comparison of UAV trajectories in ISAC analysis by revealing sensing anomalies in UAV1's path where the shaded region indicates a potential zone of instability; (b)~The scaling impact of ISAC performance in UAV networks with increasing drone numbers.}\label{jsac3}
\end{figure}

\begin{figure}[!t]
\centerline{\includegraphics[width=0.72\columnwidth]{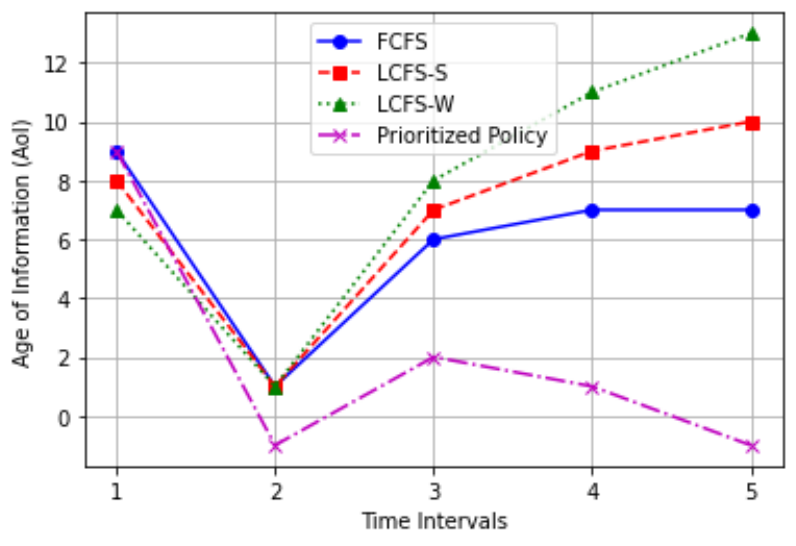}}
\vspace{-2mm}
\caption {Comparison of AoI trends in IoD network with different queuing scenarios for ISAC.}\label{aoi}
\end{figure}
\begin{figure*}[!t]
\centering
\includegraphics[width=5.8in]{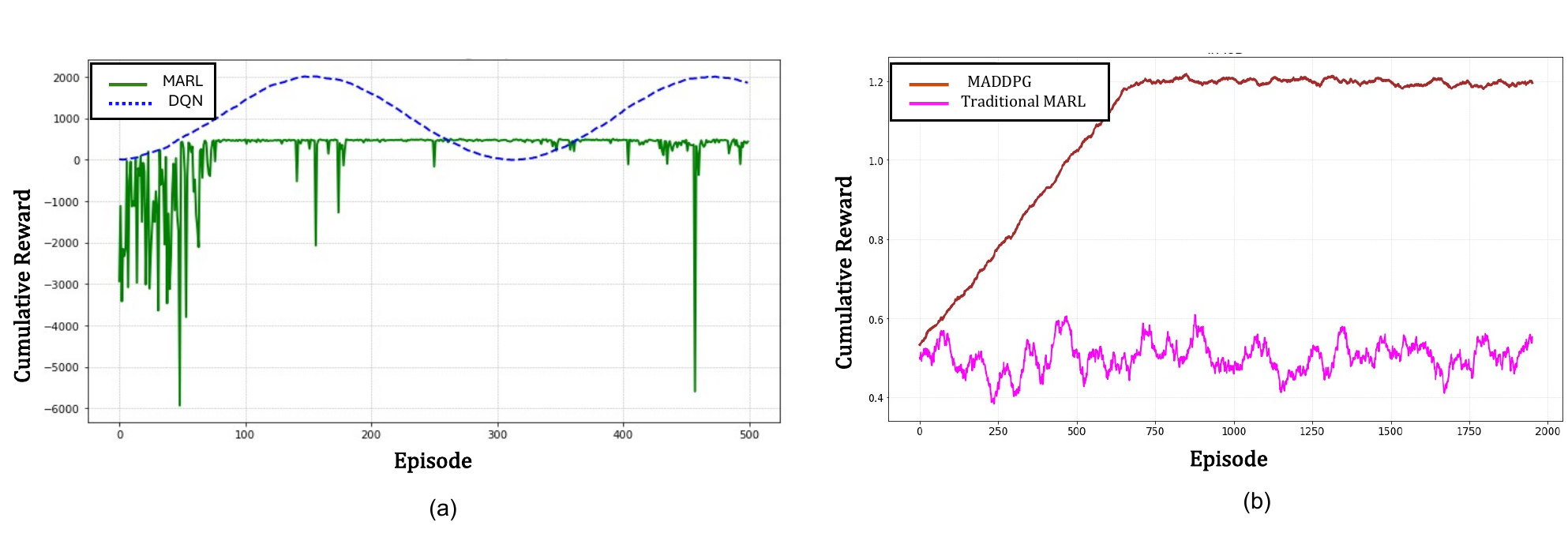}
\caption {(a) Performance Analysis on cumulative rewards when sensing is prioritized in dynamic IoD network; (b) Comparison of performance efficiency between MADDPG and traditional MARL in the IoD scenario when communication is given prominence.}\label{marl2}
%\vspace{-0.35cm}
 \end{figure*}

\textbf{Simulation Setting}: In our simulations, we consider a simple IoD network with $3$ UAVs initially and then scale up the number to $30$ UAVs. To simulate the impact of environmental conditions on sensing and communications, we have added a Gaussian noise to sensor data to reflect weather conditions, while communication signal degradation is modeled through reduced Signal-to-Noise Ratios~(SNRs) under similar adverse conditions. We have considered three primary pre-defined sensing trajectories that adjust based on real-time data and environmental conditions within the simulation area. For performing ML-based optimization, we consider the neural networks, employing three hidden layers with $128$, $256$, and $128$ neurons. We set a learning rate of $0.01$ for DQN agents and $0.001$ for MADDPG, with discount factors of $0.95$ and $0.99$, respectively. The simulations ran for $2000$ episodes, each comprising up to $200$ steps, providing a comprehensive evaluation of our proposed AdaptNet architecture in diverse and dynamic UAV operational scenarios.

This section presents visualizations and methodologies that illuminate strategies to optimize UAV operations within the IoD, focusing on using Reinforcement Learning techniques to enhance communication efficiency and significantly reduce delays. The clustering of fast-moving targets for sensing by leveraging the Fréchet distance for precise trajectory differentiation is displayed in Fig.~\ref{fig:frecblock}(a). Figure~\ref{fig:frecblock}(a) shows distinct clusters, depicted target concentrations in varying colors, while red markers with black edges represent the centroids of these clusters, signifying the average location within each group. In the visualization of Fig.~\ref{fig:frecblock}(a) for the IoD framework, UAVs dynamically determine optimal sensing paths (illustrated by blue lines) amidst clustered targets, emphasizing the role of Frechet distance across diverse regions. The functionality of the Frechet distance in directing UAVs towards efficient data sensing, clustering, and prioritization is illustrated in Figure~\ref{fig:frecblock}(b). This demonstrates a superior ability of the Frechet distance to assess the similarity in spatial-temporal data analysis, achieving a 45\% enhancement over comparative methods.
The visualization in Fig.~\ref{jsac3}(a) leverages the Fréchet distance to highlight UAV1's deviations from the expected UAV2 path. The detected anomalies within the instability region emphasize Fréchet distance's potential as a decisive metric for refining data relevance in UAV communications amidst unpredictable scenarios. Figure~\ref{jsac3}(b) illustrates how the scaling of drone numbers influences the performance of our ISAC-enabled approach. With increased drones, the sensing accuracy remains robust in maintaining operational efficiency amidst scalability.
Figure~\ref{aoi} illustrates AoI trends in joint sensing and communication for IoD networks where the curves represent diverse queuing strategies and a prioritized policy. The prioritized policy demonstrates reduced latency using Frechet distance for improved data timeliness, distinguishing it from the other queueing strategy. {In Figure~\ref{marl2}(a), MARL shows a remarkable performance advantage when sensing is prioritized, achieving convergence at 300 cumulative rewards, which signifies a convergence rate approximately 83\% faster than that of traditional Deep Q-Networks (DQN), which converge at 1800 cumulative rewards. This substantial efficiency gain shows the superiority of MARL in orchestrating collaborative UAV decision-making processes, evidencing its robust adaptability and efficiency in dynamic operational environments}. In Figure~\ref{marl2}(b), MADDPG achieves a 62\% faster convergence compared to MARL in communication-centric IoD scenarios, highlighting its efficiency in addressing communication challenges.
\section{Conclusion}
The proposed AdaptNet architecture signified a transformative leap in UAV network operations.
The strength of our AdaptNet architecture lies in its fusion of MARL's dynamic decision-making with the ISAC paradigm, enhancing UAV adaptability and precision. Moreover, implementing the MADDPG algorithm fortifies communication efficiency, setting a new standard in the IoD landscape.
 However, AdaptNet faces challenges in scalability and complexity, particularly in coordinating multiple UAVs in dense or dynamic environments, which demands high computational resources. Its reliance on advanced algorithms necessitates significant data processing capabilities, potentially hindering deployment in resource-limited scenarios. In the future, we will aim to overcome these limitations by creating more efficient algorithmic frameworks to lessen computational demands, enhancing scalability for wider deployment, and pursuing energy-saving strategies to boost UAV longevity. Additionally, integrating edge computing into AdaptNet and conducting thorough real-world testing is key to refining its practical efficacy and adaptability in diverse operational environments.
\bibliographystyle{IEEEtran}
\bibliography{main}
\begin{IEEEbiographynophoto}
{Ananya Hazarika} (Student Member, IEEE) is a Third Year PhD student at Cleveland State University, Ohio. She has completed her M. Tech from the Indian Institute of Information Technology, Guwahati. Her research interests include applications of AI/ML in wireless communication, ultra-low latency communication for extreme environments, reinforcement learning, and Bayesian optimization. She was the President of IEEE Women in Engineering (WIE) at Cleveland State University.
\end{IEEEbiographynophoto}
%\vspace{-1cm}
\begin{IEEEbiographynophoto}
{Mehdi Rahmati}
(Senior Member, IEEE) is an Assistant Professor at Cleveland State University, OH, USA, since 2020 after receiving his PhD in Electrical and Computer Engineering from Rutgers University, NJ, USA. 
He has published numerous peer-reviewed conference and journal papers and has received many prestigious awards, including the IEEE Oceanic Engineering Society Young Professional Boost award in 2022-2023,
the best demo
award at the IEEE International Conference on Sensing, Communication and Networking (SECON’19), the best paper award at the IEEE International
Conference on Mobile Ad-hoc and Sensor Systems (MASS’17), and the best
paper runner-up award at the ACM International Conference on Underwater
Networks \& Systems (WUWNet15). He was selected as the winner of the FIRST PRIZE (ex aequo) of the 2019 IEEE ComSoc Student Competition. Currently, he works on underwater and terrestrial wireless communications, connected vehicles, and Internet of Things in uncertain environments. 
\end{IEEEbiographynophoto}
%\balance

\end{document}